\documentclass[]{aa}
\usepackage{txfonts}
\usepackage{graphicx}
\usepackage{natbib}
\bibpunct{(}{)}{;}{a}{}{,} 

\newcommand{\EQ}{\begin{equation}}
\newcommand{\EN}{\end{equation}}
\newcommand{\EQA}{\begin{eqnarray}}
\newcommand{\ENA}{\end{eqnarray}}
\newcommand{\Fig}[1]{Fig.~\ref{#1}}
\newcommand{\Eq}[1]{Eq.~(\ref{#1})}
\newcommand{\Sec}[1]{Section~\ref{#1}}
\newcommand{\Tab}[1]{Table~\ref{#1}}
\newcommand{\BB}{{\bf {B}}}
\newcommand{\JJ}{{\bf {J}}}

\title{Steady state reconnection at a single 3D magnetic null point}

\author{K. Galsgaard$^1$ \and D. I. Pontin$^2$}

\institute{Niels Bohr Institute, Blegdamsvej 17, Dk-2100 Copenhagen {\O}, Denmark
\and
Division of Mathematics, University of Dundee, Dundee, DD1 4HN, United Kingdom}
\date{\today}

\abstract{}
{To systematically stress a rotationally symmetric 3D magnetic null point by advecting the opposite footpoints of the spine axis in opposite directions. This stress eventually concentrates in the vicinity of the null point forming a local current sheet through which magnetic reconnection takes place. The aim is to look for a steady state evolution of the current sheet dynamics which may provide scaling relations for various characteristic parameters of the system.}
{The evolution is followed by solving numerically the non-ideal MHD equations in a Cartesian domain. The null point is embedded in an initially constant density and temperature plasma.}
{It is shown that a quasi-steady reconnection process can be set up at a 3D null by continuous shear driving. It appears that a true steady state in unlikely to be realised as the current layer tends to grow until restricted by the geometry of the computational domain and imposed driving profile. However, ratios between characteristic quantities clearly settle after some time to stable values -- so that the evolution is quasi-steady. The experiments show a number of scaling relations, but they do not provide a clear consensus for extending to lower magnetic resistivity or faster driving velocities. More investigations are needed to fully clarify the properties of current sheets at magnetic null points.}
{}
\keywords{Magnetic fields -- Magnetohydrodynamics (MHD) -- Sun: Corona, Magnetic reconnection -- Methods: Numerical}
\authorrunning{Galsgaard \& Pontin}
\titlerunning{Steady state 3D null reconnection}

\begin{document}

\maketitle

\section{Introduction}

The solar photosphere is threaded by a complicated distribution of magnetic field concentrations of opposite polarities. Above the photosphere, in the equatorial region, the positive and negative flux concentrations typically connect to one another in a non-trivial way, creating complicated topological patterns. The simplest approach to investigate the topology of the magnetic field is to use magnetogram data as boundary conditions for potential or linear force free magnetic field extrapolations. This has been done by a number of authors \citep{2003SoPh..212..251C,2003ApJ...597L.165S,2003PhPl...10.3321L} and is found to lead to large numbers of separatrix surfaces dividing three-dimensional (3D) space into regions of different magnetic connectivity. The analysis show that 3D nulls are present in abundance at low altitudes, becoming increasingly rare high above the photosphere, and the number distribution depends on the distribution patterns of the photospheric magnetic field \citep{longcope2009}. From observations there are also a few cases in which 3D nulls have been inferred from the structure of the emission in the corona \citep{1999SoPh..185..297F}, but using this indirect evidence for a 3D null is problematic without making attempts to derive the underlying magnetic field structure. This was recently done by \cite{2009ApJ...700..559M} who verified their magnetic field extrapolations by comparing with the observed emission patterns. These independent approaches all show that 3D nulls are present in the solar atmosphere, and we must therefore investigate their role in the energy release process that takes place there.

Magnetic reconnection is though to play a leading role in a large number of different types of magnetic energy release processes where fast conversion of magnetic energy into bulk motion, plasma heating and particle acceleration is observed. These processes have been observed in a wide variety of situations, for example in ``explosive" events taking place in the outer solar atmosphere. To understand these phenomena in their various physical environments, a detailed understanding of the reconnection process is required. For many years two-dimensional (2D) reconnection models have formed the basis for this understanding. Analytical investigations have shown that various configurations with different characteristics are possible, and the driving profile controls these characteristic properties to a large extent. One key property that can be estimated is the scaling behaviour of the 2D reconnection, which informs us about how fast reconnection can take place in a given steady state situation. 

One problem that has been encountered when starting to make numerical attempts to investigate the magnetic reconnection scenario is that the imposed magnetic resistivity ($\eta$) has significant implications for the development of the reconnection region. In the case where constant resistivity is imposed, the magnetic structure has a tendency to collapse into a current sheet, but it has a difficulty in reaching a steady state situation where there is a perfect balance between the inflow and the reconnection process with a constant size current sheet. Instead, the current sheet tends to continue growing towards a solution that resembles a Sweet-Parker current sheet, with close to ``infinite" length and close to zero reconnection rate. The way to avoid the infinite current sheet is to change the $\eta$ profile. As soon as $\eta$ is made space, time or field dependent, the situation changes and the current sheet reaches a finite extent instead. 

\begin{figure}
{\hfill \includegraphics[width=0.5\textwidth]{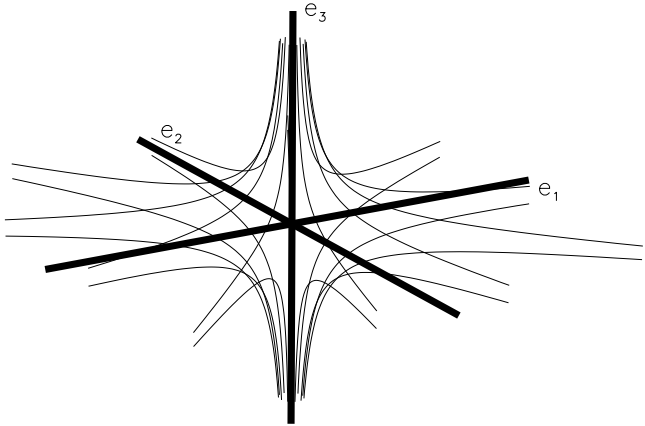} \hfill}
\caption[]{\label{single_null.fig} The structure of a single null. The $e_3$ direction represents the spine axis, while the plane defined by $e_1$ and $e_2$ represents the fan plane.}
\end{figure}

Over the last 10-15 years the focus of investigations has changed from 2D reconnection to the more challenging and realistic 3D reconnection processes. It is becoming clear that in 3D the possible reconnection scenarios are much more diverse than in 2D, and provide a series of new challenges to resolve. The process of 3D reconnection is not restricted to regions where the magnetic field vanishes as in 2D, but may also take place in regions with non-vanishing magnetic field. In this paper we will investigate in some detail the reconnection process that can take place in the vicinity of a single 3D null point. Null points constitute critical points of the magnetic field, with the only possible class being a hyperbolic critical point due to the solenoidal nature of ${\bf B}$ \citep{Green89,1990ApJ...350..672L,1996PhPl....3..759P}. A single pair of field lines approach (recede from) the null from opposite directions, and identify the so-called ``spine" of the null. A set of field lines radiate away from (approach) the null in a surface known as the ``fan" of the null (see \Fig{single_null.fig}). 

Earlier investigations have shown that reconnection at 3D nulls can take place in a number of characteristic manners. First, the topological properties of the magnetic flux evolution have been shown to be crucially dependent on the orientation of the electric current at the null \citep{2004GApFD..98..407P, 2005GApFD..99...77P}. Second, depending on the motions that drive the formation of the current layer at the null, the reconnection may occur either in a tube aligned to the spine, a planar layer in the fan, or a layer focused at the null \citep{1996ApJ...472..840R, 2007JGRA..11203103P}. The various regimes have recently been categorized by \cite{2009PhPl...16l2101P}. Due to the increased complexity of the magnetic structure, analytical solutions to the problem of single null reconnection can only be achieved using various simplifying assumptions, and no-one has at present shown how steady-state reconnection takes place at a single null in a compressible plasma, or if indeed a steady-state solution is possible. 

In the present investigation a 3D null is stressed by imposing a systematic driving velocity on the two boundaries through which the spine axis of the null penetrates. The driving is of a shear type, such that the angle between the spine and fan would be expected to change. Under the simplistic assumptions made by \cite{1996RSPSA.354.2951P} this was expected to lead to current accumulation in the entire fan plane. Although it appears that such a situation would indeed occur in an incompressible plasma \citep{1995ApJ...455L.197C, 1998PhPl....5..635C, 2007PhPl...14e2109P}, when plasma compressibility is included, a local collapse of the null occurs, destroying the planar nature of the fan surface. A localised current sheet forms which is focused around the null itself, with the result that the geometry of the magnetic field around the null is significantly different from the initial linear profile. The resulting mode of reconnection has been dubbed ``spine-fan reconnection" by \cite{2009PhPl...16l2101P} (having the appearance of a hybrid of early spine and fan models), and involves magnetic field lines being transported through/around the spine and across the fan surface.

A series of numerical experiments have investigated the reconnection process at single null. These have used various forms of driving of the system, that generally have perturbed the null region for a finite period of time \citep{2003JGRA..108.1042G,2007PhPl...14e2109P,2007PhPl...14e2106P,2007JGRA..11203103P}. This paper discusses the situation where a long time systematic driving is imposed on the system in an attempt to reach a steady state reconnection configuration. By continuously advecting the magnetic field in the vicinity of the spine axis, the magnetic field is compressed towards the null from two sides. Assuming that reconnection takes place, then due to the change in magnetic field strength away from the null, one would expect a constantly changing amount of magnetic flux to be advected towards the null per unit time. As a result, one might expect it to be difficult to obtain a true steady-state situation, while a continuously evolving quasi-steady-state situation may be more likely.

In this paper we will discuss, in \Sec{setup.sec}, the generic set-up of the problem, the numerical approach and the imposed driving of the system. In \Sec{exp.sec} the experiments will be discussed and the results highlighted. This is followed, in \Sec{discussion.sec}, by a discussion of the consequences of the results for our general understanding of the null point reconnection process. Finally, in \Sec{con.sec} we present our conclusions.

\section{Numerical setup}
\label{setup.sec}

The magnetic field configuration adopted here is the linear rotationally symmetric potential magnetic field defined by
\EQ
\BB = B_0 [-2x, y, z],
\EN
where $B_0$ is a scaling parameter, here of order unity. The null is located at $(x,y,z)=(0,0,0)$ with the spine axis along the $x$-axis and the fan plane represented by the $y-z$ plane, i.e. $x=0$. This linear null point is simple to use in numerical experiments, but represents a non-physical situation where the magnetic field strength continues to increase with distance from the null point. In reality non-linear effects will become important at some distance and change the linear growth. To investigate the dynamical evolution of the null, the domain of interest has to be limited in size. For this investigation we have limited the domain to $(x,y,z) = \pm (0.5,1.5,1.5)$, and have discretised it using a uniform grid with $200x300x300$ points.

The plasma is initially assumed to be at rest and have a constant density and thermal energy. In the present experiment the MHD equations are non-dimensionalised and the density is set to unity, while the thermal energy is set to 0.025. This means that the surface on which the plasma beta equals unity is an ellipsoid with minor axis along the $x$-axis of length 0.08 and major axis perpendicular to this with length of 0.18. Inside this surface the plasma beta tends to infinity as the null point is approached.

The linear nature of the magnetic null point forces a limitation on the numerical domain size. As a Cartesian domain does not represent very well the initial rotational symmetry of the magnetic null structure, and because we are going to impose a symmetry breaking perturbation, it is a non-trivial problem to describe the boundary conditions of the imposed domain as the perturbation reaches the boundaries. In the present approach, we force the flow perpendicular to all of the boundaries to be zero, and restrict parallel velocities at the boundaries to be non-zero only in the two regions where dedicated stressing velocity flows are imposed. 

\begin{figure}
{\hfill \includegraphics[width=0.45\textwidth]{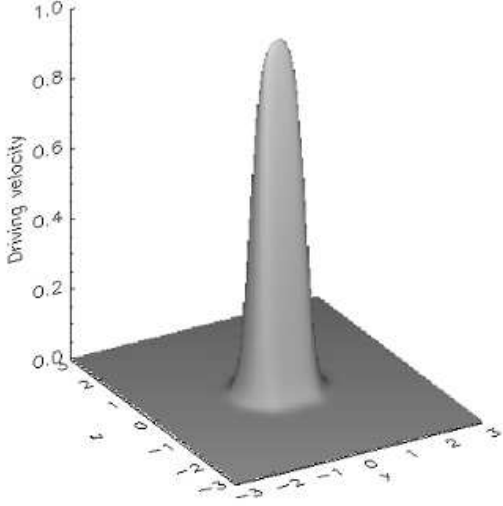} \hfill}
\caption[]{\label{driver.fig} A surface plot of the imposed driving velocity profile.}
\end{figure}
In these experiments we limit the driving of the system to two areas on the $x-$boundaries, by advecting the magnetic field in the $y$-direction, with an additional spatial dependence on the $z$-direction. The imposed driver has the following expression,
\EQA
\label{driver.eq}
V_y(y,z) & =  & \pm 0.5\,V_d\,(\tanh((y-y_0)/y_h)-\tanh((y+y_0)/y_h)) \\
         & &             ~~~~~~ \times~ (\tanh((z-z_0)/z_h)-\tanh((z+z_0)/z_h)), \nonumber
\ENA
where $V_d$ is the peak driving amplitude, and $(y,z)=(\pm y_0,\pm z_0)$ represent the locations (at the centre of the hyperbolic tangent function) where the driving velocity has half of its peak value. The variables $y_h$, $z_h$ are the half widths of the hyperbolic tangent functions. This gives a near linear driving velocity between $(y,z)=(\pm(y_0-y_h), \pm(z_0-z_h))$ while the velocity decreases outside this region, asymptotically approaching zero. For these experiments we use $y_0 = 0.6$, $y_h = 0.2$, $z_0 = 0.3$ and $z_h = 0.2$ (see \Fig{driver.fig}). This profile is imposed on the two $x$-boundaries with opposite directions.
The driving is switched on at the initiation of the experiment and reaches the peak velocity exponentially with a given time scale. 

The experiments are conducted using the 3D non-ideal MHD code by \cite{Nordlund+Galsgaard97}. This is a high order finite difference code using staggered grids to maintain conservation of physical quantities. The interpolation operators are fifth order in space while the derivative operators are sixth order. The solution is advanced in time using a third order explicit predictor-corrector method. Here we intend to derive scaling relations, so we adopt a constant $\eta$ model. 

\section{Experiments}
\label{exp.sec}
The aim of the experiments is to investigate the possibility of obtaining steady state reconnection at a 3D null point. We also seek to derive scaling relations to enable the extension of the results to higher Reynolds numbers.
In what follows we investigate the effect of varying two different global parameters of the system, namely the magnetic resistivity ($\eta$) and the imposed driving velocity ($V_d$).

An overview of the various simulation runs is given in \Tab{driver.tab}, together with three characteristic parameters for the experiments that will be discussed later. Adopting characteristic parameters for the experiment, we can determine the magnetic Reynolds numbers for these experiments $Rm={V_a L} / \eta$, where $V_a=0.5$ is the Alfv{\'e}n velocity at the spine axis at the top/bottom boundary, $L=1$ is the distance between the $x$-boundaries and $\eta$ is given in \Tab{driver.tab}. This leads to values for $Rm$ between 50 and 5000.

\begin{table}
\caption[]{\label{driver.tab} Experiments investigating the impact of the imposed boundary driving velocity, $V_d$, and the plasma resistivity, $\eta$.} 
\begin{tabular}{ l c c c c c r}
\hline
\hline
Name & $\eta$ & $V_{d}^{~~\mathrm{a}}$ & Peak $J^{~\mathrm{b}}$ & Peak $V^{~\mathrm{c}}$ & $E_{||}^{~~\mathrm{d}}$ \\ 
\hline 
A1 & $1 \cdot 10^{-2}$  & 0.100  &   8.0    & 0.75     &        \\ 
A2 & $1 \cdot 10^{-3}$  & 0.100  &  41.6    & 0.83     & 0.033  \\ 
A3 & $1 \cdot 10^{-4}$  & 0.100  & 140.0    & 0.93     &        \\ 
B  & $1 \cdot 10^{-3}$  & 0.075  &  38.4    & 0.73     & 0.026  \\ 
C1 & $1 \cdot 10^{-2}$  & 0.050  &   5.0    & 0.34     & 0.040  \\ 
C2 & $5 \cdot 10^{-3}$  & 0.050  &   9.2    & 0.42     & 0.034  \\ 
C3 & $1 \cdot 10^{-3}$  & 0.050  &  33.0    & 0.58     & 0.023  \\ 
C4 & $5 \cdot 10^{-4}$  & 0.050  &  52.0    & 0.68     & 0.019  \\ 
C5 & $1 \cdot 10^{-4}$  & 0.050  &          & 0.89     & 0.0058 \\ 
C6 & $0              $  & 0.050  &          &          &        \\ 
D  & $1 \cdot 10^{-3}$  & 0.020  &  19.3    & 0.28     & 0.013  \\ 
E  & $1 \cdot 10^{-3}$  & 0.005  &   3.5    & 0.082    & 0.0037 \\ 
\hline
\end{tabular}
\begin{list}{}{}
\item[$^{\mathrm{a}}$] imposed boundary driving velocity
\item[$^{\mathrm{b}}$] peak current value
\item[$^{\mathrm{c}}$] peak outflow velocity
\item[$^{\mathrm{d}}$] peak integrated parallel electric field
\end{list}
\end{table}

The following subsections discuss different aspects of the experiments. We begin with a general description of the dynamical evolution, first discussing a 2D setup as a reference for the 3D experiment, since our knowledge of the behaviour in the 2D case is well established (at least within the MHD framework).

\subsection{Evolution of a 2D system}\label{2dsec}

\begin{figure}
{\hfill \includegraphics[width=0.5\textwidth]{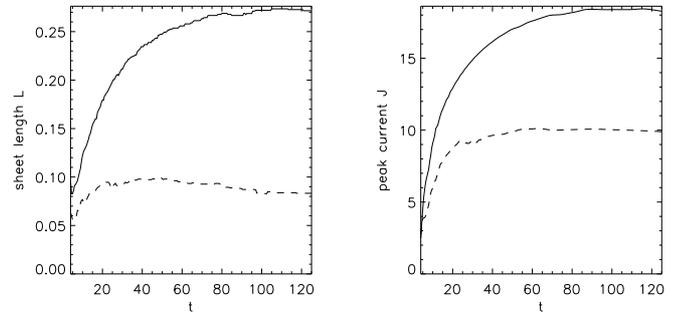} \hfill}
\caption[]{\label{width_current_2D.fig} Shearing of a 2D X-point. Left: sheet length as a function of time for domain  $[x,y] \in [\pm 0.5, \pm 1.5]$ (dashed line) and $[x,y] \in [\pm 1, \pm 3]$ (solid line). Right: peak current as a function of time for the same two domain sizes.}
\end{figure}

We first briefly describe the evolution of a sheared 2D null point. The magnetic field is taken to be ${\bf B}=[-x,y,0]$ and the $z$-dependence of the driver is suppressed. As the driving is switched on two MHD waves propagate towards the null point from either of the driving boundaries. Close to the null they collide and a strong current layer forms in the region where the two separatrices collapse very close to one another. The length of the current sheet and the maximum current density (and therefore in 2D the reconnection rate) gradually build to peak values, before beginning a slow decay (see \Fig{width_current_2D.fig}). During this period, the configuration can be described as being in a quasi-steady state.

There are a number of key points to note. First, the halt in the growth of the current layer (both length and modulus), is due to the finite size of our computational domain. If we repeat the simulations in a larger computational domain, with an extended driving region, the current layer continues to grow for substantially longer, before again being constrained by the boundaries (compare the full and dashed lines in \Fig{width_current_2D.fig}). This is a well-known effect in 2D -- that for uniform resistivity a `Sweet-Parker' type current layer develops, which continually grows as the forcing continues. The limitation on the current layer in these simulations is therefore equivalent to the limitation imposed when reconnection is forced in an initial Harris sheet with periodic boundaries \citep[e.g.][]{2005GeoRL..3206105B}. Another important point to note is that, were the driving really constant in our simulations, we would not expect such a noticeable decay of the current layer length and modulus after the peak is reached. The decay is due to a reduction in the quantity of flux being forced into the sheet per unit time as the simulations progress. This in turn is due to the evacuation of flux from the region upstream of the inflow, as flux from that region is constantly fed through the sheet. The above properties all have analogues in the 3D system, as discussed below.

\subsection{Qualitative evolution of the 3D system}\label{3d.sec} 
\begin{figure}
{\hfill \includegraphics[width=0.5\textwidth]{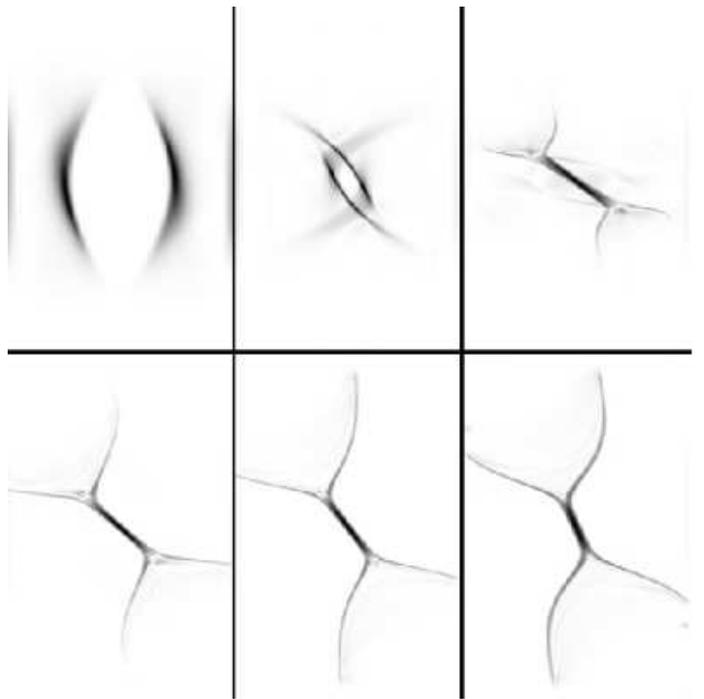} \hfill}
\caption[]{\label{current_movie.fig} Current amplitude at a number of characteristic instances in the dynamical evolution. The individual frames are scaled to their local dynamical range to enhance the visual appearance. The images represent the $z=0$ plane for run ``C3" at times $t=(0.4, 1.0, 2.2, 4.4, 9.2, 19.0)$.}
\end{figure}

Initiation of the driving on the two $x$-boundaries launches two wavefronts that propagate through the magnetised plasma. As the driving speed is slowly ramped up the wavefronts initially propagate towards the fan plane as simple plane Alfv{\'e}n waves (with their $x$ speed being independent of $y$ and $z$). With the driving continuously stressing the magnetic field, the field lines behind the wavefront are systematically advected in the $y$-direction. This breaks the initial symmetry of the field, generating an asymmetry of the field on either side of the advected spine line. This causes the wavefront to tilt as its speed becomes dependent on $y$, due to an increase in wave speed in the region where the field lines are stretched, \Fig{current_movie.fig}. 

From the initial phase of the evolution, two points should be noted. First, the width of the current front decreases while its intensity increases on its way towards the fan plane. This is an effect of the decreasing Alfv{\'e}n velocity in the $x$-direction, that for a pure Alfv{\'e}n wave implies that it can never reach the fan plane of the null point in a strict mathematical sense. This results in a pileup producing a strong current front. As this occurs, the wavefront also spreads out increasingly fast along the $y$- and $z$-directions as it approaches the fan plane, due to the field geometry. One would expect this to act to reduce the current intensity in the front, were this effect the dominant one. However, due to the compression of the plasma, the wave is not a pure Alfv{\'e}n wave, but couples to both slow and fast mode waves. The perturbation can therefore propagate across the magnetic field lines and advance towards the null point. This type of behaviour was seen in \cite{2003JGRA..108.1042G}, where the null point was perturbed by two rotational twist waves centred on the spine axis. Here the fast mode wave wraps around the null and concentrates part of the energy of the perturbation there \citep[see also][who studied the 2D case]{2009A&A...493..227M}. 

As the evolution proceeds, the two wavefronts initiated at opposite $x$-boundaries approach each other and eventually collide, with the two non-rotationally-symmetric current concentrations compressing into one central current distribution centred on the null. Within the resulting current layer, the spine and fan have collapsed towards one another, and the current spreads along the fan both along and across the driving direction. The tilt and magnitude of the arising current concentration depend on the imposed driving speed and imposed $\eta$ as will be discussed later. Typically the tilt of the current sheet overshoots initially and the sheet experiences a few oscillations before settling to a stable angle. Due to the imposed constant $\eta$, reconnection takes place in the current sheet, and a near steady-state balance between the flux advected into the current sheet and the reconnected flux exiting the sheet through a reconnection jet is reached (see below). This picture is maintained with only minor changes until the termination of the experiments.

\subsection{Steady state evolution}\label{steadysec}
\begin{figure}
{\hfill \includegraphics[width=0.5\textwidth]{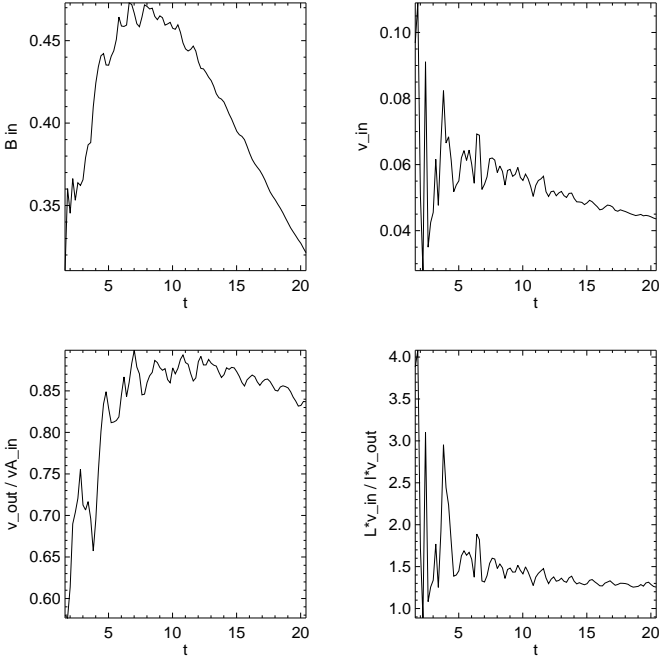} \hfill}
\caption[]{\label{steady-state.fig} The time evolution of the magnetic field in the inflow region (top left), the plasma inflow velocity (top right), the outflow velocity relative to the Alfv{\'e}n velocity in the inflow region and the ratio between the inflow (lower left) and outflow velocities and sheet dimensions defined by $L\,v_{in}/(l\,v_{out})$ (lower right). For the run C3.}
\end{figure}
\begin{figure}
{\hfill \includegraphics[width=0.5\textwidth]{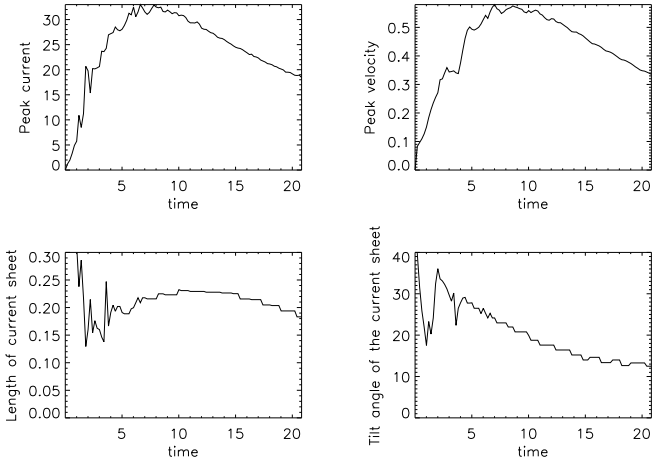} \hfill}
\caption[]{\label{time_dependence.fig} The four panels show the time evolution of the peak value of the current in the sheet (top left), the peak velocity of the reconnection jet (top right), the length of the current sheet (lower left) and finally its angle relative to the $y$-axis (lower right), for the run C3.}
\end{figure}

From the 2D experiment it is seen that the evolution reaches a quasi-steady-state. In order for it to be physically meaningful to search for scaling relations, we must first check if such a quasi-steady-state also exists in the 3D simulations. To determine this we examine some of the characteristic parameters addressed in the Sweet-Parker \citep{1957JGR....62..509P,1958IAUS....6..123S} 2D reconnection model. 

In order to define the characteristic size of the current layer we define the boundary of the layer to be the surface at which the current modulus $|{\bf J}|$ falls to 50\% of its peak value (this peak occurring at the null). In 3D this provides three characteristic length scales that are here listed in descending order; The {\it length} ($L$) in the plane of the shear perturbation (here the $xy$-plane), the extent of the sheet in the direction perpendicular to this plane and parallel to the current at the null (here the $z$ direction) which we denote here by the {\it width} ($w$), and finally the {\it thickness}, $l$, of the sheet (measured in the $z=0$ plane). 

First we estimate the amount of magnetic flux being transported into the sheet -- for which one needs to know both the inflow velocity perpendicular to the current sheet and the magnetic vector aligned with the sheet. For the purposes of making such an estimate, we measure these values at the centre of the inflow region just outside the current sheet. We also consider the outflow jet velocity, measured toward the outflow edge of the current sheet. \Fig{steady-state.fig} shows four combinations of these variables as functions of time from experiment ``C3". The top left frame represents the component of the magnetic field parallel to the current sheet. It shows an initial build-up followed by a near linear decline in magnitude. The latter is a consequence of the depletion of the magnetic field in the inflow region due to the limited length scale of the driving in the $y$-direction. 
The top right panel shows the inflow velocity, which is found to reach a stable level with only a slow decay with time. The lower left frame shows the ratio of the outflow velocity and the Alfv{\'e}n speed in the inflow region (akin to the reconnection rate in the Sweet-Parker model), which reaches a nearly stable value at the same time as the magnetic field component reached its peak value. Finally the lower right frame shows the ratio of the length scale of the inflow region times the inflow velocity and the thickness of the outflow region times the outflow velocity. In the 2D Sweet Parker analysis this relation constitutes mass conservation, and here we can see that it reaches a stable value declining only slowly with time. It is interesting to note that this stable value is somewhat greater than 1 (the 2D value), being closer to a value around 1.5. This is likely due to the fact that the outflowing plasma may access the third dimension in 3D, rather than being constrained to flow out in the plane of the inflow.

In \Fig{time_dependence.fig} the top left panel shows the time development of the maximum current density in the domain, which reaches a peak around $t=7$ followed by a slow decay. The top right panel represents the time development of the jet velocity, which is slightly different from the current, taking longer to reach its maximum value. This is expected since the velocity is a collective effect of the Lorentz tension force and gas pressure in the newly reconnected field lines, acting on a ``dense" plasma and therefore needing some additional time to accelerate the plasma to it cruising velocity. The lower left panel shows how the sheet length is very stable in time, while the lower right panel represents the tilt angle of the sheet that shows a slow decreases with time. 

From the above discussion it appears that the effect of the continual driving is to force the null structure into a near steady state, in which the reconnection region maintains its characteristics for a period of time much longer than the Alfv{\'e}n travel time along the current sheet (\Fig{steady-state.fig} and \Fig{time_dependence.fig}). The qualitative behaviour discussed above is also observed in the other experiments listed in \Tab{driver.tab} above. We therefore go on in the following section to derive scaling relations between various parameters in the simulations. 

\subsection{Scaling relations}\label{scalingsec}

\subsubsection{Peak current}
\begin{figure}
{\hfill \includegraphics[width=0.45\textwidth]{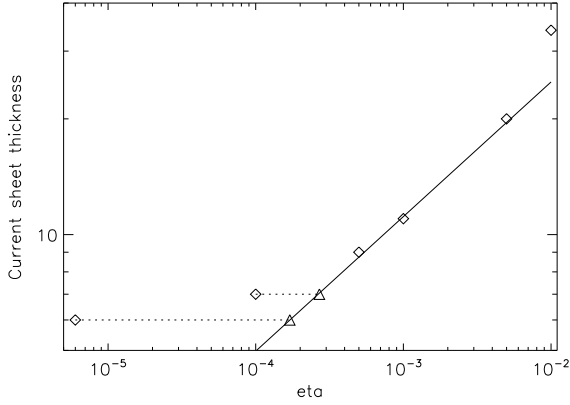} \hfill}
\caption[]{\label{width_scale.fig} Scaling relation between the imposed $\eta$ and the thickness of the current sheet (measured in units of the grid spacing in the $x$-direction), for the runs $C1-C6$. The full line represents sheet thickness $\propto \eta^{0.35}$.}
\end{figure}
\begin{figure}
{\hfill \includegraphics[width=0.45\textwidth]{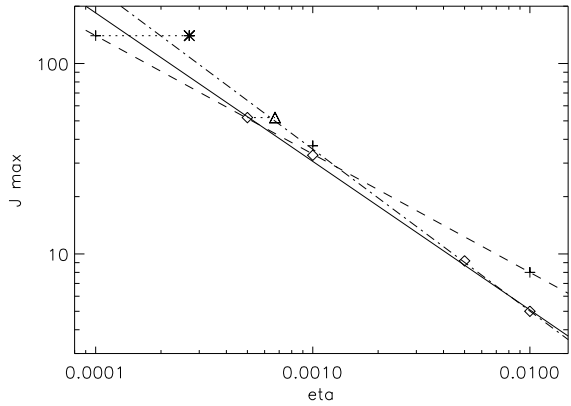} \hfill}
\caption[]{\label{current_scaling_eta.fig} The scaling relation of the peak current with the imposed $\eta$ value. The data points represent two series of experiments with different imposed driving velocities. The diamonds and the associated full line represent data with $V_d$=0.05. The triangle represents the corrected $\eta$ for the left most point as indicated in the previous graph, with the dot-dashed line being the correspondingly altered scaling. The crosses and the dashed line represent the data with $V_d$=0.1 (with the star showing the corrected $\eta$ value). The dashed line represents Eq. 3, the full line Eq. 4 and the dot-dashed line Eq. 5.}
\end{figure}

It is well known that the peak current $J_{max}$ reached in a numerical experiment depends on two key parameters, one being the numerical resolution. If the imposed $\eta$ is too low then the numerical diffusion dominates and will limit $J_{max}$. The other key parameter is the chosen value of the resistivity $\eta$, which limits the current growth through the ``correct" evolution of the induction equation. With sufficient numerical resolution one can resolve the structure of the current sheet and it becomes possible to make a meaningful analysis of the impact of changing $\eta$. In this investigation we have run simulations with a limited selection of $\eta$ values.

One way to investigate whether the imposed resistivity dominates over the numerical resistivity is to measure the width of the current sheet as a function of the imposed $\eta$ value. In the incompressible case \cite{2004SoPh..222...95H} showed that this gives rise to a power law dependence with a power of 0.5 (although we note that the morphology of the current sheet is somewhat different in their incompressible models to what is observed here). \Fig{width_scale.fig} shows the relation for the ``C" experiments given in \Tab{driver.tab}. The full line is a power law, but with a power of 0.35 for the higher $\eta$ values. In the plot the diamonds represent the imposed values of $\eta$, with the point furthest to the left representing a case where $\eta$ was set to zero. This indicates that the code introduces an efficient $\eta$ when the width of the sheet becomes on the order of the numerical stencil. If one assumes that the power law dependence can be extended, then an asymptotic value of the numerical $\eta$ can be estimated for the given resolution. Correcting the values of $\eta$ correspondingly for the runs $C5$ and $C6$ (two left most points on the graph) brings those data points onto the line representing the scaling relation (diamonds in the plot). In the following we will investigate if this correction of the imposed resistivity will have any implication on the scaling relations.

When data from the experiments listed in \Tab{driver.tab} are used, scaling relations of the peak current with $\eta$ can be obtained. \Fig{current_scaling_eta.fig} shows plots of peak current versus $\eta$ for two different driving velocities. The triangle represents the ``corrected" $\eta$ value for the run ``C4" and the star is the corrected value for ``A3". The scalings, corresponding to the lines plotted in the Figure, are 
\EQA
\label{I_scaling_eta1.eq}
J_{max} & = & 0.46\, \eta^{-0.62} \qquad \mbox{ for }V_d=0.1, \\ 
\label{I_scaling_eta4.eq}
J_{max} & = & 0.14\, \eta^{-0.78} \qquad \mbox{ for }V_d=0.05 -  \mbox{without $\eta$ correction,} \\
J_{max} & = & 0.1\, \eta^{-0.85}  \qquad \, \,\, \mbox{ for }V_d=0.05 - \mbox{with $\eta$ correction.}
\ENA
 Here one can see two different comparable fits for the slow ($V_d=0.05$) driving experiments, with the scaling based on the corrected value leading to a slightly stronger dependence of $J_{max}$ on $\eta$.
However, the correction of the $\eta$ value for the faster driver does not seem to lead to a linear scaling relation. 

\begin{figure}
{\hfill \includegraphics[width=0.45\textwidth]{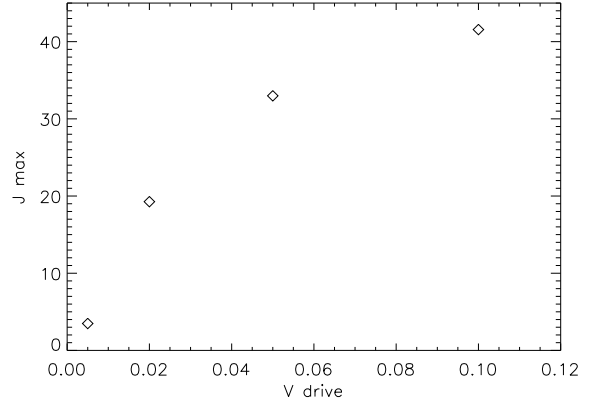} \hfill}
\caption[]{\label{current_scaling_vd.fig} The scaling relation between the imposed driving velocity and the peak current for the experiments with $\eta= 10^{-3}$ listed in \Tab{driver.tab}.}
\end{figure}

We now turn to the scaling of the peak current with the driving velocity, for a constant $\eta$ value, which is investigated using the $\eta=10^{-3}$ experiments listed in \Tab{driver.tab}. The result is shown in \Fig{current_scaling_vd.fig}, where it is seen that while the peak current clearly depends on the imposed driving velocity, there is no simple (e.g.~linear or exponential) relation that can be obtained to fit the data. 

\subsubsection{Current layer dimensions and jet velocity}

The dimensions and tilt angle of the current sheet have been measured for the same experiments. They are found to be independent of $\eta$ for a constant driving velocity, to within our measurement accuracy. On the other hand, varying the driving velocity while maintaining a constant value for $\eta$, we find that there are significant changes of these dimensions, indicating that they are largely dependent on the amount of imposed stress. However, as discussed above, the present domain size and the extent of the driving region limit the current sheet from evolving freely, and thus with the present setup we are unable to provide scaling relations for these quantities. 

\begin{figure}
{\hfill \includegraphics[width=0.45\textwidth]{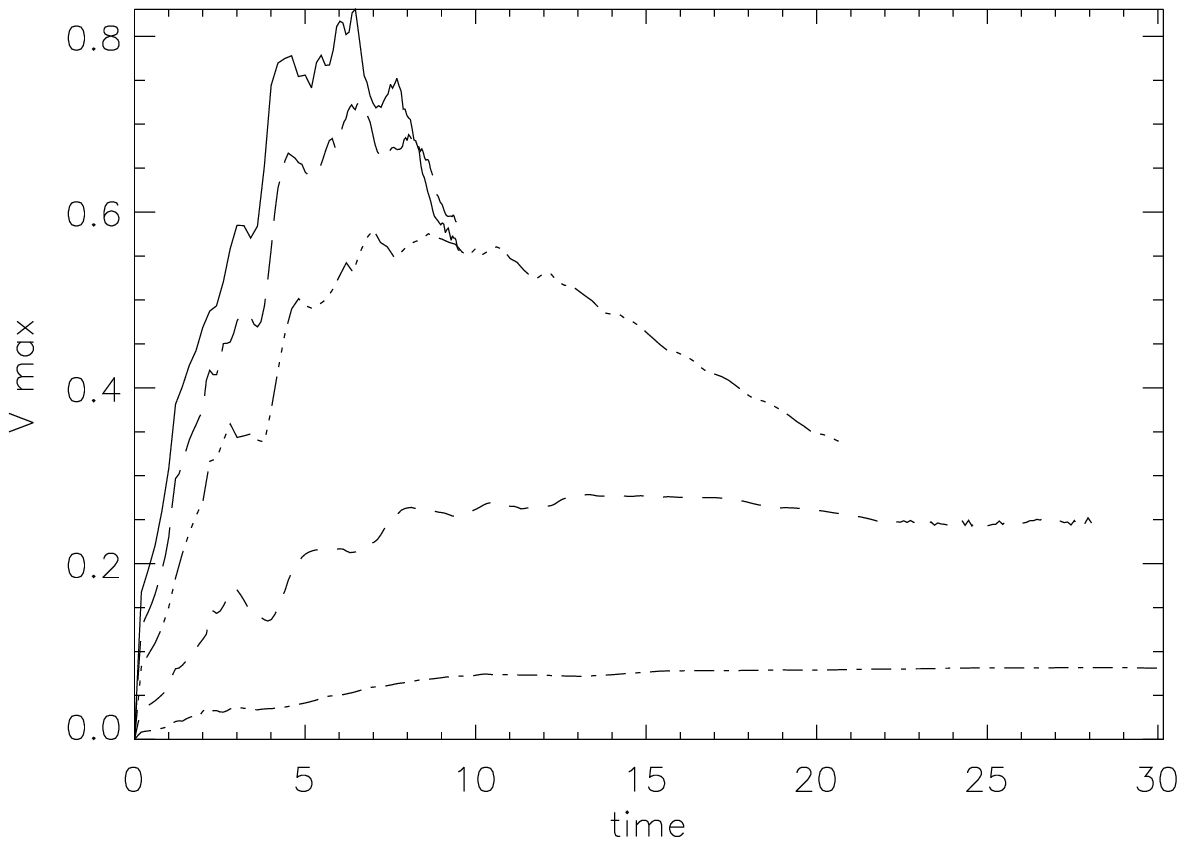} \hfill}

{\hfill \includegraphics[width=0.45\textwidth]{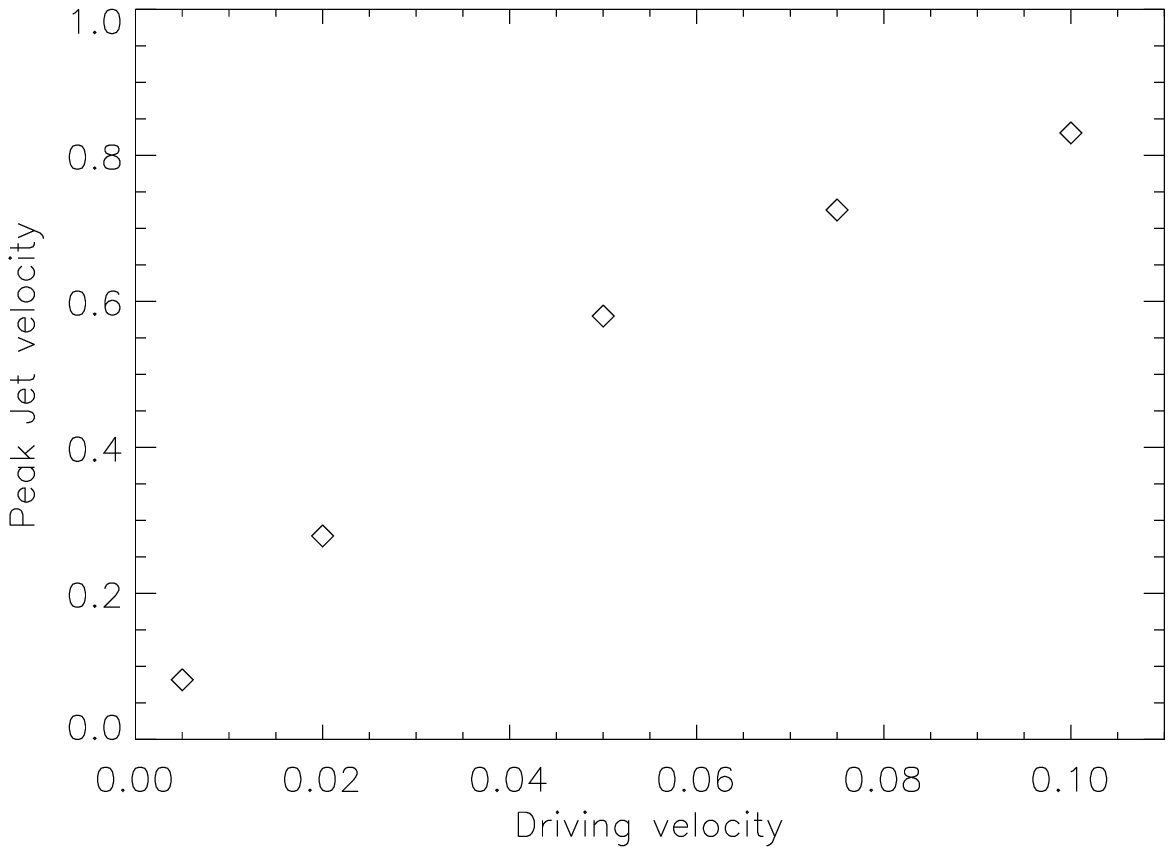} \hfill}

{\hfill \includegraphics[width=0.45\textwidth]{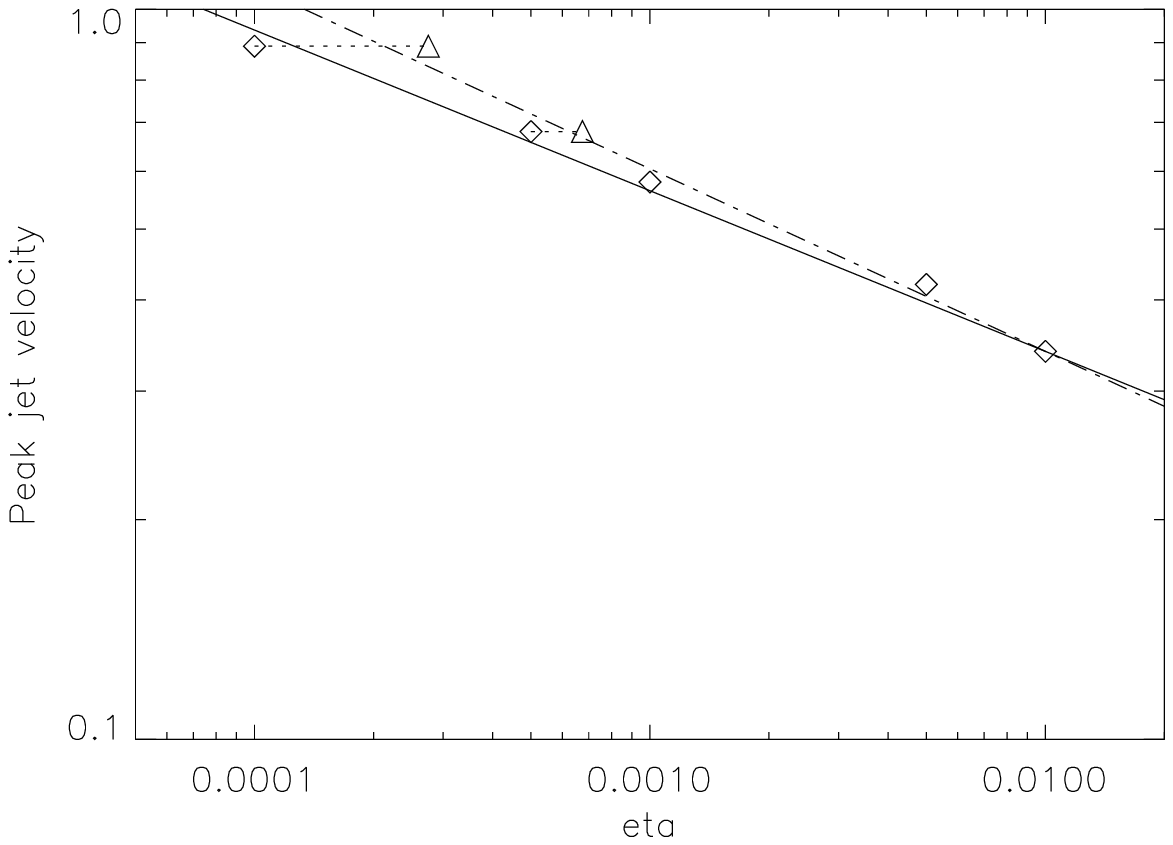} \hfill}
\caption[]{\label{velocity_scaling.fig} The top frame shows the time dependent evolution of the peak velocity, for the runs C1 (dot-dashed), C2 (dashed), C3 (triple-dot-dashed), C4 (long-dashed) and C5 (solid) (see \Tab{driver.tab}). The middle frame shows the peak velocity as a function of the imposed driving velocity. The lower frame shows the peak velocity as a function of the imposed resistivity, $\eta$. The full line represents the scaling using the original values, {\bf Eq.(6)}, while the dot-dashed line included the corrected $\eta$ values{\bf, Eq.(7)}.}
\end{figure}

The top frame of \Fig{velocity_scaling.fig} shows the maximum outflow velocity as a function of time for different driving velocities $V_d$. This shows a significant variation both with time and imposed driving speed. The middle frame shows the peak outflow jet velocity (for all time) as a function of the imposed driving velocity. Here we can see that we have a non-linear dependence, which is also not fit well with an exponential scaling, but follows a similar pattern to the scaling of $J_{max}$ with $V_d$. Finally, the lower frame of \Fig{velocity_scaling.fig} shows the jet velocity as a function of the imposed resistivity, where we find approximately power law scalings, with
\EQA
V_{peak} = {1 \over 8.1}\eta ^{-0.22} -  \mbox{without $\eta$ correction,}\\
V_{peak} = {1 \over 9.3}\eta ^{-0.25} - \mbox {with $\eta$ correction}.
\ENA
The first scaling is for the case where we use the manually applied $\eta$ values for the experiments, while in the second the corrected values are included.

\subsection{Reconnection rate}
\label{rec_rate.sec}
From established theory \citep{1988JGR....93.5547S,2003JGRA..108.1285P} we know that the correct measurement of the reconnection rate in 3D is given by the maximal value of $\Phi$, the integrated electric field parallel to a magnetic field line ($E_\|$), where this maximum is taken over all field lines which thread the non-ideal region (assumed to be spatially localised). To determine the reconnection rate we therefore perform a systematic field line tracing for a collection of field lines that thread the current sheet, and seek the maximum in the integrated parallel electric field. In practice this occurs along a field line that is located very close to the $z$-axis (it would be expected to lie exactly along the $z$-axis based on the exactly symmetric analytical model of \cite{2005GApFD..99...77P}). In these experiments a constant $\eta$ has been used, which implies that $E_\|=\eta J_\|$, so that
\EQ
\label{e_par.eq}
\Phi = \int_\epsilon^L (\eta \JJ \cdot \BB)/|\BB| dl,
\EN
where we integrate along the given field line starting close ($\epsilon$) to the null out to a finite distance, $L$. To get the correct reconnection rate the integral has to be made in the $z$-direction in both senses, to $\pm L$. In previous simulations \cite[e.g.][]{2007PhPl...14e2106P} this was a relatively simple process, as the diffusion region was limited to a finite size. However, the situation here is different. Due to the continual driving, the current not only has a strong component close to the null, but there also exists an extended (though weaker) concentration of current in the fan plane that reaches all the way to the $z$-boundaries. 
\begin{figure}
{\hfill \includegraphics[width=0.45\textwidth]{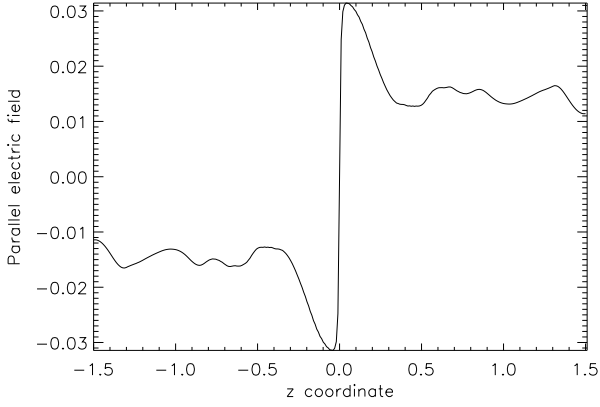} \hfill}

{\hfill \includegraphics[width=0.45\textwidth]{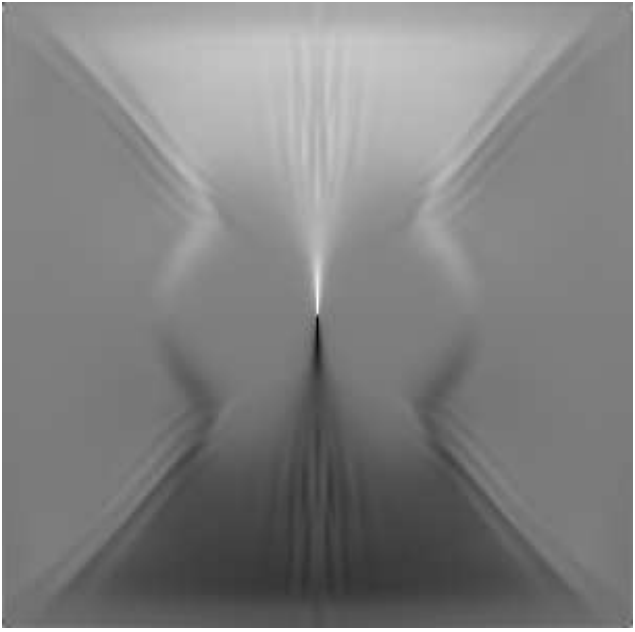} \hfill}
\caption[]{\label{e_par.fig} The top panel shows the parallel electric field ($E_\|={\bf E}\cdot{\bf B}/|{\bf B}|$) along the $z$-axis for the run ``C3". The bottom panel shows a 2D image of the parallel electric field in the $x=0$ plane.}
\end{figure}
\begin{figure}
{\hfill \includegraphics[width=0.49\textwidth]{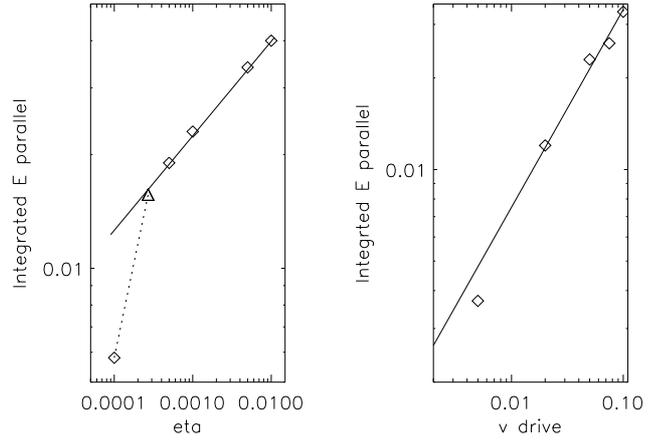} \hfill}
\caption[]{\label{e_par_scale.fig} The left panel shows the dependence of the reconnection rate on the imposed $\eta$ with a constant driving velocity ($V_d=0.05$), with the triangle showing the corrected value for the run ``C5" as discussed above{\bf, Eq.(9)}. The right panel shows the dependence on the reconnection rate on the imposed driving velocity, for constant $\eta$ ($\eta=10^{-3}$){\bf, Eq.(10)}.}
\end{figure}
This is demonstrated in the top image in \Fig{e_par.fig}, which shows a 1D plot of the parallel electric field along the $z$-axis (which is not a magnetic field line). In the lower frame a 2D $(y,z)$ representation of the same quantity is presented, which again shows the extended parallel electric field. Plotting the same quantity in a plane perpendicular to this shows that the parallel electric field is by contrast limited to a very narrow layer in the $x$-direction (compare with \Fig{current_movie.fig}). The actual reconnection process is therefore not limited to the local region around the strong current sheet, but extends all the way to the imposed boundary. This clearly implies that the value obtained using \Eq{e_par.eq} depends on the choice of $L$. For the results below the value of $L$ has been limited to $L=1.4$, such that the domain boundary is not reached. 

Performing the integration defined in \Eq{e_par.eq} on the various data sets, we obtain the results shown in \Fig{e_par_scale.fig}. The left frame shows the peak value of $\Phi$ as a function of $\eta$. If we correct the result obtained with the lowest $\eta$ value (run $C5$, left most diamond in the plot) relative to the results obtained from \Fig{width_scale.fig}, this point is shifted to the location marked by the triangle in the plot, that falls nicely on the straight line scaling relation that fits the other data points. This simple correction seems to work here, most likely because the current sheet in the diffusion region is all the way just being resolved with a minimum number of gridpoints. 

In the right hand frame, an indication of a scaling relation between the driving velocity and reconnection rate can be seen, although with some fluctuations around the obtained line. The two power law relations are given by:
\EQA
\label{eta_scale.eq}
\Phi & =& \int E_{\|} dl = 0.126 ~ \eta^{0.25} ~~~~~ {\rm for} ~~V_d=0.05 \\ 
\Phi & =& \int E_{\|} dl = 0.15 ~ V_d^{0.65} ~~~~~~  {\rm for} ~~ \eta=10^{-3}
\ENA

\section{Discussion}
\label{discussion.sec}

The experiments described above were conducted to investigate whether a steady-state magnetic reconnection process could be set up in 3D by continually stressing a symmetric null point. The results presented in \Sec{steadysec} show that a quasi-steady state exists where the inflow to the diffusion region is balanced by the ongoing magnetic reconnection (this is true for all values of the imposed driving velocity and $\eta$). 
Due to the magnetic geometry, the inflow magnetic field and the plasma inflow and outflow velocities ($v_{in}$ and $v_{out}$) are not constant during this quasi-steady evolution. However, ratios between characteristic quantities familiar from 2D models do settle to approximately constant values. For example, the ratio of $v_{out}$ to the inflow Alfv{\' e}n speed is approximately fixed. So is the ratio $L v_{in}/l v_{out}$, which characterises mass conservation in the sheet and settles to a constant value of around 1.5 (rather than 1 as in 2D, since the outflowing plasma may access the third dimension rather than being constrained to flow out in the plane of the inflow).
It should be noted that in each of the simulations the quasi-steady configuration obtained is strongly influenced by the restricted dimensions of both the numerical domain and the region on which the boundary driving is imposed. If it was possible to extend the domain and driving region indefinitely, indications are that the current layer would extend indefinitely, until reaching a threshold for some instability such as the tearing instability.


We have investigated the dependence of different characteristic quantities in the quasi-steady state on both the driving velocity $V_d$ and imposed resistivity $\eta$. Considering first the dependence on $V_d$, we found that the current maximum  and outflow velocity increase as $V_d$ increases, although not with a simple scaling relation. It is natural that these quantities increase when the rate at which flux is forced into the sheet is increased, and also that they behave in the same way since the outflow jet is accelerated by the ${\bf J}\times{\bf B}$ force.

For the simulation runs with different values of $\eta$ we have been able to determine approximate scaling relationships. The peak current appears to scale like $J_{max}\sim \eta^{-0.6}-\eta^{-0.8}$, which is consistent with the $\eta^{-3/4}$ scaling expected for planar incompressible fan current layers \citep{1998PhPl....5..635C}. The peak outflow velocity on the other hand follows an approximate scaling $V_{peak}\sim\eta^{-0.25}$, which mirrors the scaling of the reconnection rate, as we discuss below.
It is also clear from \Fig{width_scale.fig} that the width of the current sheet depends on the imposed $\eta$ value, and the graph indicates that as $\eta$ approaches zero the sheet reaches a minimum thickness, controlled by numerical diffusion. Results for large $\eta$ indicate a power law dependence for the thickness as a function of $\eta$, though with a smaller exponent than found in the incompressible models \citep{1998PhPl....5..635C,2004SoPh..222...95H}. One can use this to estimate the value of the numerical diffusivity, which takes effect only when structures collapse to the resolution limit, while its value decreases rapidly as the structures increase in size. Making use of this information to correct the scaling parameters in different plots is not simple, as indicated by the additional points included in a number of figures. In some cases it seems to provide a scaling relation, while in other cases, it makes no coherent correction to the result. The only way to provide better information for these values of $\eta$ is through experiments with even higher numerical resolution. However, doubling the numerical resolution locally makes the experiments much more expensive. To make a serious difference in the scaling at least a factor of 10 variation in resolution is required -- which gives a factor of $10^4$ in required computing time. It is therefore a significant computational challenge to improve these scaling relations significantly for lower $\eta$ values. 

From 2D reconnection theory the simple Sweet-Parker  model gives a scaling of the reconnection rate that depends on $\eta^{0.5}$. With realistic parameters for astrophysics plasmas this reconnection rate is far too slow to make it an attractive mechanism for magnetic energy conversion, in for instance solar flares. It has therefore since then been a task to find reconnection setups that depends less strongly on $\eta$ and therefore evolve much faster. Taking the scaling rate for the reconnection rate given in \Eq{eta_scale.eq}, the driven fan-spine reconnection is found to scale with a smaller power of $\eta$, namely 0.25. 
This scaling for the reconnection rate  is significantly faster than the slow Sweet-Parker scaling, though slower than the fast Petcheck reconnection rate that scales with the logarithm of $\eta$. However, it is worth pointing out that over our range of $\eta$ covering two orders of magnitude ($\eta=10^{-2}-10^{-4}$) it is difficult to distinguish the power law and $\ln(\eta)$ scalings.

The obtained scaling relations should be compared with those obtained for the same setup, but with an impulsive (i.e.~temporally localised) perturbation of the null point, described by \cite{2007PhPl...14e2106P} and \cite{2009PhPl...16l2101P}. In these investigations, the boundary driving was applied for a fixed period of time and then reduced to zero. It is worth noting that this implies that the net displacement of the spine -- under the assumption of an ideal evolution -- increases as the driving speed increases. Their analysis showed that for fixed $\eta$, the peak current and reconnection rate increase linearly with the driving velocity, while the length and width of the sheet ($L$ and $w$ above) actually decrease linearly. This seems counter-intuitive, but is a result of the fact that the collapse of the field in the vicinity of the null is more efficient for stronger driving. Thus, the scaling of the current and reconnection rate follow similar patterns to those seen above in the continually driven case, however the sheet dimensions scale in the opposite way. Turning now to the scaling with resistivity $\eta$ (for fixed driving speed) in the impulsively driven case, it was found that as $\eta$ decreases, the peak current increases and the reconnection rate decreases, with a rate between power law and logarithmic, while here we may argue in favor for a power law dependence. Clearly there are some similarities, but also some differences, between the two different driving configurations.

\section{Conclusions}
\label{con.sec}
The experiments have shown that it is possible for a 3D null point to reach a quasi-steady state, where the characteristic parameters of the current sheet are controlled by a combination of the imposed driving velocity and the effective constant resistivity. It is likely that other parameters, such as the dimensions of the plasma-$\beta=1$ surface also play a role. It is found that scaling relations between characteristic parameters do not always follow simple exponential or power law expressions over the full range of $\eta$ values. This is partly due to the limited numerical resolution in the simulations, which means that the code imposes its own diffusion when structures becomes too small. Only with substantially higher numerical resolution will it be possible to push the reliable range of $\eta$ to significantly smaller values. However, this preliminary study indicates that the reconnection rate scales relatively weakly with the plasma resistivity, as $\eta^{1/4}$.

The scaling relations found here are different from the ones derived for the impulsive perturbation of the single null. However, we find that even changing the amplitude of the fixed driving velocity changes the scaling with resistivity. It is therefore not easy to make any predictions regarding the global behaviour of 3D nulls in realistic situations: too many parameters seem to influence the dynamical evolution, with the scalings being linked in ways that appear to be non-linear, and that we are therefore yet to unravel. Further to this, test experiments indicate that some of the obtained parameters are likely to change when the size of the imposed driver is changed.

It is clear from the results that more effort has to be invested in understanding null reconnection and its scaling behaviour before we are able to make any quantitative predictions regarding processes occurring at 3D nulls in the Solar atmosphere or the Earth's magnetosphere.

\begin{acknowledgements}
Thanks to the anonymous referee for making suggestions to view some of the results in a different light. 
Computing time was given by the DCSC at university of Copenhagen. Support by the European Commission through the Solaire Network (MTRNCT-2006-035484) is gratefully acknowledged. DP acknowledges support from the Royal Society.
\end{acknowledgements}

\bibliographystyle{aa} 
\bibliography{ref} 
\end{document}